\documentclass[aps,prl,superscriptaddress,showpacs,floatfix,twocolumn]{revtex4}
\usepackage{graphicx}   % Include figure files
\usepackage{epsfig}
\usepackage{dcolumn}
\usepackage{bm}
\usepackage{amssymb}

\newcommand{\pt}{\mbox{$p_\mathrm{T}\,$}}

\begin{document}

\title{Transverse momentum and centrality dependence of dihadron correlations 
       in Au+Au collisions at $\sqrt{s_{\rm{NN}}}$ = 200 GeV: Jet-quenching 
       and the response of partonic matter}

\newcommand{\abilene}{Abilene Christian University, Abilene, TX 79699, U.S.}
\newcommand{\banaras}{Department of Physics, Banaras Hindu University, Varanasi 221005, India}
\newcommand{\bnl}{Brookhaven National Laboratory, Upton, NY 11973-5000, U.S.}
\newcommand{\caucr}{University of California - Riverside, Riverside, CA 92521, U.S.}
\newcommand{\charlesczech}{Charles University, Ovocn\'{y} trh 5, Praha 1, 116 36, Prague, Czech Republic}
\newcommand{\ciae}{China Institute of Atomic Energy (CIAE), Beijing, People's Republic of China}
\newcommand{\cns}{Center for Nuclear Study, Graduate School of Science, University of Tokyo, 7-3-1 Hongo, Bunkyo, Tokyo 113-0033, Japan}
\newcommand{\colorado}{University of Colorado, Boulder, CO 80309, U.S.}
\newcommand{\columbia}{Columbia University, New York, NY 10027 and Nevis Laboratories, Irvington, NY 10533, U.S.}
\newcommand{\czechtech}{Czech Technical University, Zikova 4, 166 36 Prague 6, Czech Republic}
\newcommand{\dapnia}{Dapnia, CEA Saclay, F-91191, Gif-sur-Yvette, France}
\newcommand{\debrecen}{Debrecen University, H-4010 Debrecen, Egyetem t{\'e}r 1, Hungary}
\newcommand{\elte}{ELTE, E{\"o}tv{\"o}s Lor{\'a}nd University, H - 1117 Budapest, P{\'a}zm{\'a}ny P. s. 1/A, Hungary}
\newcommand{\fit}{Florida Institute of Technology, Melbourne, FL 32901, U.S.}
\newcommand{\fsu}{Florida State University, Tallahassee, FL 32306, U.S.}
\newcommand{\gsu}{Georgia State University, Atlanta, GA 30303, U.S.}
\newcommand{\hiroshima}{Hiroshima University, Kagamiyama, Higashi-Hiroshima 739-8526, Japan}
\newcommand{\ihepprot}{IHEP Protvino, State Research Center of Russian Federation, Institute for High Energy Physics, Protvino, 142281, Russia}
\newcommand{\illuiuc}{University of Illinois at Urbana-Champaign, Urbana, IL 61801, U.S.}
\newcommand{\instpasczech}{Institute of Physics, Academy of Sciences of the Czech Republic, Na Slovance 2, 182 21 Prague 8, Czech Republic}
\newcommand{\isu}{Iowa State University, Ames, IA 50011, U.S.}
\newcommand{\jinrdubna}{Joint Institute for Nuclear Research, 141980 Dubna, Moscow Region, Russia}
\newcommand{\kaeri}{KAERI, Cyclotron Application Laboratory, Seoul, South Korea}
\newcommand{\kek}{KEK, High Energy Accelerator Research Organization, Tsukuba, Ibaraki 305-0801, Japan}
\newcommand{\kfki}{KFKI Research Institute for Particle and Nuclear Physics of the Hungarian Academy of Sciences (MTA KFKI RMKI), H-1525 Budapest 114, POBox 49, Budapest, Hungary}
\newcommand{\korea}{Korea University, Seoul, 136-701, Korea}
\newcommand{\kurchatov}{Russian Research Center ``Kurchatov Institute", Moscow, Russia}
\newcommand{\kyoto}{Kyoto University, Kyoto 606-8502, Japan}
\newcommand{\labllr}{Laboratoire Leprince-Ringuet, Ecole Polytechnique, CNRS-IN2P3, Route de Saclay, F-91128, Palaiseau, France}
\newcommand{\lawllnl}{Lawrence Livermore National Laboratory, Livermore, CA 94550, U.S.}
\newcommand{\losalamos}{Los Alamos National Laboratory, Los Alamos, NM 87545, U.S.}
\newcommand{\lpc}{LPC, Universit{\'e} Blaise Pascal, CNRS-IN2P3, Clermont-Fd, 63177 Aubiere Cedex, France}
\newcommand{\lund}{Department of Physics, Lund University, Box 118, SE-221 00 Lund, Sweden}
\newcommand{\muenster}{Institut f\"ur Kernphysik, University of Muenster, D-48149 Muenster, Germany}
\newcommand{\myongji}{Myongji University, Yongin, Kyonggido 449-728, Korea}
\newcommand{\nagasaki}{Nagasaki Institute of Applied Science, Nagasaki-shi, Nagasaki 851-0193, Japan}
\newcommand{\newmex}{University of New Mexico, Albuquerque, NM 87131, U.S. }
\newcommand{\nmsu}{New Mexico State University, Las Cruces, NM 88003, U.S.}
\newcommand{\ornl}{Oak Ridge National Laboratory, Oak Ridge, TN 37831, U.S.}
\newcommand{\orsay}{IPN-Orsay, Universite Paris Sud, CNRS-IN2P3, BP1, F-91406, Orsay, France}
\newcommand{\peking}{Peking University, Beijing, People's Republic of China}
\newcommand{\pnpi}{PNPI, Petersburg Nuclear Physics Institute, Gatchina, Leningrad region, 188300, Russia}
\newcommand{\riken}{RIKEN, The Institute of Physical and Chemical Research, Wako, Saitama 351-0198, Japan}
\newcommand{\rikjrbrc}{RIKEN BNL Research Center, Brookhaven National Laboratory, Upton, NY 11973-5000, U.S.}
\newcommand{\rikkyo}{Physics Department, Rikkyo University, 3-34-1 Nishi-Ikebukuro, Toshima, Tokyo 171-8501, Japan}
\newcommand{\saispbstu}{Saint Petersburg State Polytechnic University, St. Petersburg, Russia}
\newcommand{\saopaulo}{Universidade de S{\~a}o Paulo, Instituto de F\'{\i}sica, Caixa Postal 66318, S{\~a}o Paulo CEP05315-970, Brazil}
\newcommand{\seoulnat}{System Electronics Laboratory, Seoul National University, Seoul, South Korea}
\newcommand{\stonybrkc}{Chemistry Department, Stony Brook University, Stony Brook, SUNY, NY 11794-3400, U.S.}
\newcommand{\stonycrkp}{Department of Physics and Astronomy, Stony Brook University, SUNY, Stony Brook, NY 11794, U.S.}
\newcommand{\subatech}{SUBATECH (Ecole des Mines de Nantes, CNRS-IN2P3, Universit{\'e} de Nantes) BP 20722 - 44307, Nantes, France}
\newcommand{\tenn}{University of Tennessee, Knoxville, TN 37996, U.S.}
\newcommand{\titech}{Department of Physics, Tokyo Institute of Technology, Oh-okayama, Meguro, Tokyo 152-8551, Japan}
\newcommand{\tsukuba}{Institute of Physics, University of Tsukuba, Tsukuba, Ibaraki 305, Japan}
\newcommand{\vandy}{Vanderbilt University, Nashville, TN 37235, U.S.}
\newcommand{\waseda}{Waseda University, Advanced Research Institute for Science and Engineering, 17 Kikui-cho, Shinjuku-ku, Tokyo 162-0044, Japan}
\newcommand{\weizmann}{Weizmann Institute, Rehovot 76100, Israel}
\newcommand{\yonsei}{Yonsei University, IPAP, Seoul 120-749, Korea}
\affiliation{\abilene}
\affiliation{\banaras}
\affiliation{\bnl}
\affiliation{\caucr}
\affiliation{\charlesczech}
\affiliation{\ciae}
\affiliation{\cns}
\affiliation{\colorado}
\affiliation{\columbia}
\affiliation{\czechtech}
\affiliation{\dapnia}
\affiliation{\debrecen}
\affiliation{\elte}
\affiliation{\fit}
\affiliation{\fsu}
\affiliation{\gsu}
\affiliation{\hiroshima}
\affiliation{\ihepprot}
\affiliation{\illuiuc}
\affiliation{\instpasczech}
\affiliation{\isu}
\affiliation{\jinrdubna}
\affiliation{\kaeri}
\affiliation{\kek}
\affiliation{\kfki}
\affiliation{\korea}
\affiliation{\kurchatov}
\affiliation{\kyoto}
\affiliation{\labllr}
\affiliation{\lawllnl}
\affiliation{\losalamos}
\affiliation{\lpc}
\affiliation{\lund}
\affiliation{\muenster}
\affiliation{\myongji}
\affiliation{\nagasaki}
\affiliation{\newmex}
\affiliation{\nmsu}
\affiliation{\ornl}
\affiliation{\orsay}
\affiliation{\peking}
\affiliation{\pnpi}
\affiliation{\riken}
\affiliation{\rikjrbrc}
\affiliation{\rikkyo}
\affiliation{\saispbstu}
\affiliation{\saopaulo}
\affiliation{\seoulnat}
\affiliation{\stonybrkc}
\affiliation{\stonycrkp}
\affiliation{\subatech}
\affiliation{\tenn}
\affiliation{\titech}
\affiliation{\tsukuba}
\affiliation{\vandy}
\affiliation{\waseda}
\affiliation{\weizmann}
\affiliation{\yonsei}
\author{A.~Adare}	\affiliation{\colorado}
\author{S.~Afanasiev}	\affiliation{\jinrdubna}
\author{C.~Aidala}	\affiliation{\columbia}
\author{N.N.~Ajitanand}	\affiliation{\stonybrkc}
\author{Y.~Akiba}	\affiliation{\riken} \affiliation{\rikjrbrc}
\author{H.~Al-Bataineh}	\affiliation{\nmsu}
\author{J.~Alexander}	\affiliation{\stonybrkc}
\author{A.~Al-Jamel}	\affiliation{\nmsu}
\author{K.~Aoki}	\affiliation{\kyoto} \affiliation{\riken}
\author{L.~Aphecetche}	\affiliation{\subatech}
% ...
\collaboration{PHENIX Collaboration} \noaffiliation

\date{\today}

\begin{abstract}
Azimuthal angle ($\Delta\phi$) correlations are presented for
charged hadrons from dijets for $0.4<\pt<10$~GeV/$c$ in Au+Au
collisions at $\sqrt{s_{\rm{NN}}}$ = 200 GeV. With increasing 
\pt, the away-side distribution evolves from a broad to a
concave shape, then to a convex shape. Comparisons to $p+p$ data
suggest that the away-side can be divided into a partially
suppressed ``head'' region centered at $\Delta\phi\sim\pi$, and an
enhanced ``shoulder'' region centered at $\Delta\phi\sim
\pi\pm1.1$. The \pt spectrum for the ``head'' region softens
toward central collisions, consistent with the onset of jet
quenching. The spectral slope for the ``shoulder'' region is
independent of centrality and trigger \pt, which offers
constraints on energy transport mechanisms and suggests that the
``shoulder'' region contains the medium response to energetic jets.
\end{abstract}

\pacs{25.75.Dw}

\maketitle

High transverse momentum (\pt) partons are valuable probes of
the high energy density matter created at the Relativistic
Heavy-Ion Collider (RHIC). These partons lose a large fraction of
their energy in the matter prior to forming hadrons, a phenomenon
known as jet-quenching. Such energy loss is predicted to lead
to strong suppression of both single- and correlated away-side
dihadron yields at high \pt~\cite{Gyulassy:2003mc}, consistent
with experimental findings~\cite{Adler:2003au,Adler:2002tq}.
The exact mechanism for energy loss is not yet understood. Recent
results of dihadron azimuthal angle ($\Delta\phi$) correlations
have indicated strong modification of the away-side
jet~\cite{Adams:2005ph,Adler:2005ee,Adler:2002tq,Adare:2006nr}.
For high \pt hadron pairs, such modification is manifested by a
partially suppressed away-side peak at
$\Delta\phi\sim\pi$~\cite{Adler:2002tq}. This has been interpreted
as evidence for the fragmentation of jets that survive their
passage through the medium.

For intermediate \pt charged hadron pairs, the away-side jet was
observed to peak at
$\Delta\phi\sim\pi\pm1.1$~\cite{Adler:2005ee,Adare:2006nr},
suggesting that the energy lost by high \pt partons is
transported to lower \pt hadrons at angles away from
$\Delta\phi\sim\pi$. The proposed mechanisms for such energy
transport include medium deflection of hard~\cite{Chiu:2006pu} or
shower partons~\cite{Armesto:2004pt},large-angle gluon
radiation~\cite{Vitev:2005yg,Polosa:2006hb}, Cherenkov gluon
radiation~\cite{Koch:2005sx}, and ``Mach Shock'' medium
excitations~\cite{Casalderrey-Solana:2004qm}.

In this letter we present a detailed ``mapping'' of the \pt and
centrality dependence of away-side jet shapes and yields. These
measurements (1) allow a detailed investigation of the jet
distributions centered around $\Delta\phi \sim \pi\pm1.1$ and
$\Delta\phi \sim \pi$, (2) provide new insight on the interplay
between jet quenching and the response of the medium to the lost
energy, and (3) provide new constraints for distinguishing the
competing mechanisms for energy transport.

The results presented here are based on minimum-bias (MB) Au+Au
and p+p datasets as well as a ``photon'' level-1 triggered (PT)
p+p dataset~\cite{Adare:2006hc} collected with the PHENIX
detector~\cite{Adcox:2003zm} at $\sqrt{s_{\rm{NN}}}$=200 GeV,
during the 2004-2005 RHIC running periods. The collision vertex
$z$ was required to be within $|z|<$~30cm of the nominal crossing
point. The event centrality was determined via the method in
Ref.~\cite{Adcox:2003zm}. A total of 840 million Au+Au events were
analyzed. Charged particles were reconstructed in the two central
arms of PHENIX, each covering -0.35 to 0.35 in pseudo-rapidity and
$90^{\circ}$ in azimuth. The tracking system consists of the drift
chambers and two layers of multi-wire proportional chambers with
pad readout (PC1 and PC3), achieving a momentum resolution of
$0.7\% \bigoplus 1.1\%~p$~(GeV/$c$)~\cite{Adler:2003au}.

Dihadron azimuthal angle correlations are obtained by correlating
``trigger'' (type $A$) hadrons with ``partner'' (type $B$) hadrons.
The MB and PT p+p datasets are used for trigger $\pt<5$ GeV/$c$
and $\pt>5$ GeV/$c$, respectively. To reduce background from decays
and conversions, tracks are required to have a matching hit within
a $\pm 2.3 \sigma$ window in PC3. For $\pt>4$ GeV/$c$, additional
matching hit at the electromagnetic calorimeter (EMC) was required
to suppress background tracks that randomly associate with the
PC3~\cite{Adler:2003au}. For triggers with $\pt>5$ GeV/$c$, a
\pt dependent energy cut in the EMC and a tight $\pm1.5\sigma$
matching cut at the PC3 were applied to reduce the background to
$<10$\%~\cite{Adler:2005ad}. This energy cut greatly reduces PT
trigger bias effects. The PT p+p results are consistent with the
MB p+p data for trigger $\pt>5$ GeV/$c$.

The jet associated partner yield per trigger,
$Y_{\rm{jet}}\left(\Delta\phi\right)$, is obtained from the
$\Delta\phi$ correlations as ~\cite{Adler:2005ad,Adler:2005ee}:
\small\begin{eqnarray} \label{eq:jet} Y_{\rm{jet}} =
\left[\frac{N^{\rm{s}}\left(\Delta\phi\right)}{N^{\rm{m}}\left(\Delta\phi\right)}-
b_0\left(1+2v_2^{A} v_2^{B} \cos2\Delta
\phi\right)\right]\frac{\int d\Delta\phi
N^{\rm{m}}(\Delta\phi)}{2\pi N_{A}\varepsilon_B}
\end{eqnarray}\normalsize
where $N_{A}$ is the number of triggers, $\varepsilon_B$ is the
single particle efficiency for partners in the full azimuth and
$\left|\eta\right|<0.35$; $N^{\rm{s}}(\Delta\phi)$ and
$N^{\rm{m}}(\Delta\phi)$ are pair distributions from the same- and
mixed-events, respectively. Mixed-event pairs are obtained by
selecting partners from different events with similar centrality
and vertex. The $\varepsilon_B$ values include detector acceptance
and reconstruction efficiency, with an uncertainty of $\sim
10$\%~\cite{Adler:2005in,Adler:2003au}. The harmonic term,
$2v_2^{A} v_2^{B} \cos2\Delta \phi$, reflects the elliptic flow
modulation of the combinatoric pairs in Au+Au
collisions~\cite{Adler:2005ee}. Values for $v_2^A$ and $v_2^B$ for
each centrality class are measured via the reaction plane (RP)
method~\cite{Adler:2003kt} using the Beam-Beam Counters at $3
<|\eta|<4$. The systematic errors on $v_2$ are dominated by the
RP resolution, and are estimated to be $\sim 6$\% for central and
mid-central collisions, and $\sim 10$\% for the peripheral
collisions~\cite{Adler:2005ee}.

\begin{figure}[tb]
\includegraphics[width=1.0\linewidth]{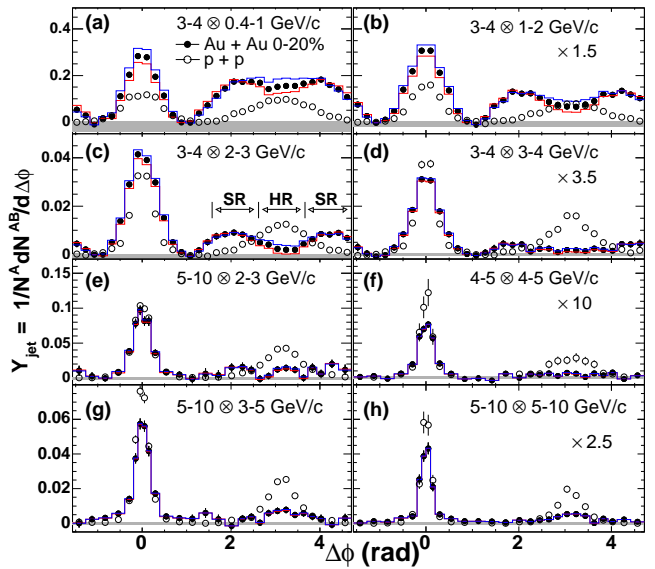}
\caption{\label{fig:shape} Per-trigger yield versus $\Delta\phi$ for
various trigger and partner $p_{T}$ ($\pt^{A}\otimes \pt^{B}$),
arranged by increasing pair momentum (sum of $\pt^{A}$ and
$p_{T}^{B}$), in $p+p$ and 0-20\% Au+Au collisions. The Data in
some panels are scaled as indicated. Solid lines (shaded bands)
indicate elliptic flow (ZYAM) uncertainties. Arrows in (c)
indicate ``head'' ({\bf HR}) and ``shoulder'' ({\bf SR}) regions.}
\end{figure}

To fix the value of $b_0$, we followed the subtraction procedure
of Refs.~\cite{Adler:2005ee,Ajitanand:2005jj} and assumed that
$Y_{\rm{jet}}$ has zero yield at its minimum $\Delta \phi_{min}$
(ZYAM). To estimate the possible over-subtraction at
$\Delta\phi_{min}$, we calculate $b_0$ values independently by
fitting $Y_{\rm{jet}}(\Delta\phi)$ to a function consisting of one
near-side and two symmetric away-side Gaussians. The fitting
procedure is similar to that used in~\cite{Adare:2006nr}, except
that a region around $\pi$ ($|\Delta\phi-\pi|<1$) is excluded to
avoid ``punch-through'' jets around $\pi$ (see
Fig.\ref{fig:shape}). This fit accounts for the overlap of the
near- and away-side Gaussians at $\Delta\phi_{min}$, and thus
gives systematically lower $b_0$ values than that for ZYAM. We
assign the differences as one-sided systematic errors on $b_0$.
This over-subtraction error is only significant in central
collisions and at $\pt^{A,B}<3$ GeV/$c$.

The per-trigger yield distributions for $p+p$ and 0-20\% central
Au+Au collisions are compared in Fig.~\ref{fig:shape} for various
combinations of trigger and partner $p_{T}$ ranges ($\pt^A\otimes
\pt^B$) as indicated. The $p+p$ data show essentially Gaussian
away-side peaks centered at $\Delta\phi\sim\pi$ for all $\pt^{A}$
and $\pt^{B}$. In contrast, the Au+Au data show substantial shape
modifications dependent on $\pt^{A}$ and $\pt^{B}$. For a fixed
value of $\pt^{A}$, Figs.~\ref{fig:shape}(a)-(d)
reveal a striking evolution from a broad, roughly flat peak to a
local minimum at $\Delta\phi\sim\pi$ with side-peaks at
$\Delta\phi \sim\pi\pm1.1$. Interestingly, the location of the
side-peaks in $\Delta\phi$ is roughly constant with increasing
$\pt^{B}$ (see also~\cite{Adare:2006nr}). Such \pt independence
is compatible with the away-side jet modification expected from a
medium-induced ``Mach Shock''~\cite{Casalderrey-Solana:2004qm} but
disfavors models which incorporate large angle gluon
radiation~\cite{Vitev:2005yg,Polosa:2006hb}, Cherenkov gluon
radiation~\cite{Koch:2005sx} or deflected
jets~\cite{Armesto:2004pt,Chiu:2006pu}.

For relatively high values of $\pt^{A}\otimes \pt^{B}$,
Figs.~\ref{fig:shape}(e)-(h) show that the away-side
jet shape for Au+Au gradually becomes peaked as for $p+p$, albeit
suppressed. This ``re-appearance'' of the away-side peak seems due
to a reduction of the yield centered at $\Delta\phi \sim\pi\pm1.1$
relative to that at $\Delta\phi\sim \pi$, rather than a merging of
the peaks centered at $\Delta\phi \sim\pi\pm1.1$. This is
consistent with the dominance of dijet fragmentation at large
$\pt^{A}\otimes \pt^{B}$, possibly due to jets that
``punch-through'' the medium~\cite{Renk:2006pk} or those emitted
tangentially to the medium's surface~\cite{Loizides:2006cs}.

The evolution of the away-side jet shape with \pt (cf.
Fig.~\ref{fig:shape}) suggests separate contributions from a
medium-induced component centered at $\Delta\phi \sim \pi\pm1.1$
and a fragmentation component centered at $\Delta\phi \sim \pi$. A
model independent study of these contributions can be made by
dividing the away-side jet function into equal-sized ``head''
($|\Delta\phi-\pi|<\pi/6$, HR) and ``shoulder''
($\pi/6<|\Delta\phi-\pi|<\pi/2$, SR) regions, as indicated in
Fig.~\ref{fig:shape}(c). We characterize the relative amplitude of
these two regions with the ratio, $R_{\rm{HS}}$,
\begin{eqnarray}
R_{\rm{HS}}  = {{\frac{{\int_{\Delta\phi\in\rm{HR}} {d\Delta \phi
Y_{\rm{jet}} (\Delta \phi )} }} {\rm{\Delta\phi_{HR}}}}
\mathord{\left/
 {\vphantom {{\frac{{\int_{\rm{HR}} {d\Delta \phi Y_{\rm{jet}} (\Delta \phi )} }}
{\rm{HR}}} {\frac{{\int_{\rm{SR}} {d\Delta \phi Y_{\rm{jet}}
(\Delta \phi )} }} {\rm{SR}}}}} \right.
 \kern-\nulldelimiterspace} {\frac{{\int_{\Delta\phi\in\rm{SR}} {d\Delta \phi Y_{\rm{jet}} (\Delta \phi )} }}
{\rm{\Delta\phi_{SR}}}}}
\end{eqnarray}
Since $N_{A}$ in Eq.\ref{eq:jet} cancels in the ratio,
$R_{\rm{HS}}$ is a pure pair variable and is symmetric $w.r.t$
$p_{T}^A$ and $p_{T}^B$:
$R_{\rm{HS}}(p_{T}^A,p_{T}^B)=R_{\rm{HS}}(p_{T}^B,p_{T}^A)$. For
concave and convex shapes, one expects $R_{\rm{HS}}<1$ and
$R_{\rm{HS}}> 1$, respectively.

\begin{figure}[tb]
\includegraphics[width=1.0\linewidth]{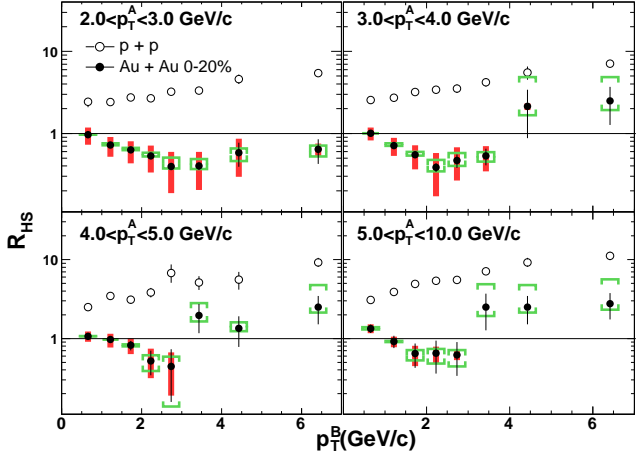}
\caption{\label{fig:ratio} 
$R_{\rm{HS}}$ versus $\pt^B$ for $p+p$
(open) and Au+Au (filled) collisions for four trigger selections.
Since $R_{\rm{HS}}$ is purely hadron pair variable, the result is
unchanged by swapping $\pt^A$ and $\pt^B$. Shaded bars (brackets)
represent \pt-correlated uncertainties due to elliptic flow
(ZYAM procedure).}
\end{figure}

Figure \ref{fig:ratio} summarizes the $p_{T}^B$ dependence of
$R_{\rm{HS}}$ for both $p+p$ and central Au+Au collisions in four
$p_{T}^A$ bins. The ratios for $p+p$ are always above one and
increase with $p_{T}^{\rm{B}}$. This reflects the narrowing of a
peaked jet shape with increasing
$\pt^{\rm{B}}$~\cite{Adler:2005ad}. In contrast, the ratios for
Au+Au show a non-monotonic dependence on $\pt^{A,B}$. They
evolve from $R_{\rm{HS}} \sim 1$ for $p_{T}^{A,B}\lesssim1$
GeV/$c$ through $R_{\rm{HS}}< 1$ for $1\alt p_{T}^{A,B}\alt 4$
GeV/$c$ followed by $R_{\rm{HS}}> 1$ for $p_{T}^{A,B} \agt 5 $
GeV/$c$. These trends reflect the competition between
medium-induced modification and jet fragmentation, and suggest
that the latter dominates at $p_{T}^{A,B} \agt 5 $ GeV/$c$.
The results shown in Fig.~\ref{fig:shape} indicate that, relative
to $p+p$, the Au+Au yield is suppressed in the HR but is enhanced
in the SR. We quantify this suppression/enhancement via
$I_{\rm{AA}}$, the ratio of jet yield $Y_{jet}$ between Au+Au and
$p+p$ collisions over a $\Delta\phi$ region, W, $I_{\rm{AA}}^W  =
{{\int_{\Delta\phi\in W} {d\Delta \phi Y_{\rm{jet}}^{\rm{Au + Au}}
} } \mathord{\left/
 {\vphantom {{\int_{\Delta\phi\in W} {d\Delta \phi Y_{\rm{jet} }^{\rm{Au + Au}} } } {\int_{\Delta\phi\in W} {d\Delta \phi Y_{\rm{jet} }^{p + p} } }}} \right.
 \kern-\nulldelimiterspace} {\int_{\Delta\phi\in W} {d\Delta \phi Y_{\rm{jet} }^{p + p} } }}$. 

\begin{figure}[tb]
\includegraphics[width=1.0\linewidth]{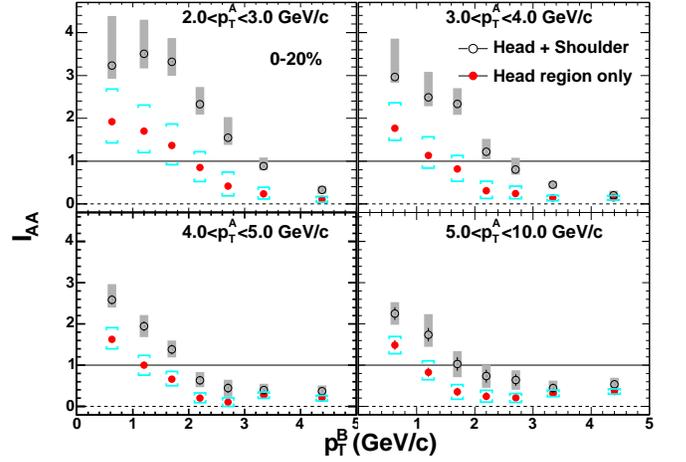}
\caption{\label{fig:inte} $I_{\rm{AA}}$ versus $\pt^B$ for four
trigger \pt bins in HR+SR ($|\Delta\phi-\pi|<\pi/2$) and HR
($|\Delta\phi-\pi|<\pi/6$). The total systematic errors for the
two regions, represented by shaded bars and brackets respectively,
are strongly correlated. Grey bands around $I_{\rm{AA}}=1$
represent 14\% combined uncertainty on the single particle
efficiency in Au+Au and $p+p$.}
\end{figure}

Figure~\ref{fig:inte} shows $I_{\rm{AA}}$ as a function of
$\pt^B$ for the HR and the HR+SR, respectively, in four
$\pt^{A}$ bins. For triggers of $2<\pt^A<3$ GeV/$c$,
$I_{\rm{AA}}$ for HR+SR exceeds one at low $\pt^B$, but falls and
crosses one at $\sim$3.5 GeV/$c$. A similar trend is observed for
the higher \pt triggers, but the enhancement (at low $\pt^B$) is
smaller and the suppression (at high $\pt^B$) is stronger. The
$I_{\rm{AA}}$ values in HR are lower relative to HR+SR for all
$\pt^{A,B}$. For the low \pt triggers, the suppression sets in
around $1\alt \pt^B \alt 3$ GeV/$c$, followed by a fall-off for
$\pt^B \agt 4$ GeV/$c$. For higher \pt triggers, a constant
level of $\sim 0.2-0.3$ is observed above $\sim 2$ GeV/$c$ similar
to the suppression level of inclusive hadrons~\cite{Adler:2003au}.
These results provide clear evidence for significant yield
enhancement in the SR and suppression in the HR. The data suggest
that the SR reflects the dissipative processes that redistribute
the energy lost in the medium; The suppression for the HR is
consistent with jet quenching. However, we note that the
$I_{\rm{AA}}$ values for the HR are upper limit estimates for the
jet fragmentation component. This is because the HR yield includes
possible contributions from the tails of the SR, as well as from
bremsstrahlung gluon radiations~\cite{Vitev:2005yg}.

\begin{figure}[tb]
\includegraphics[width=1.0\linewidth]{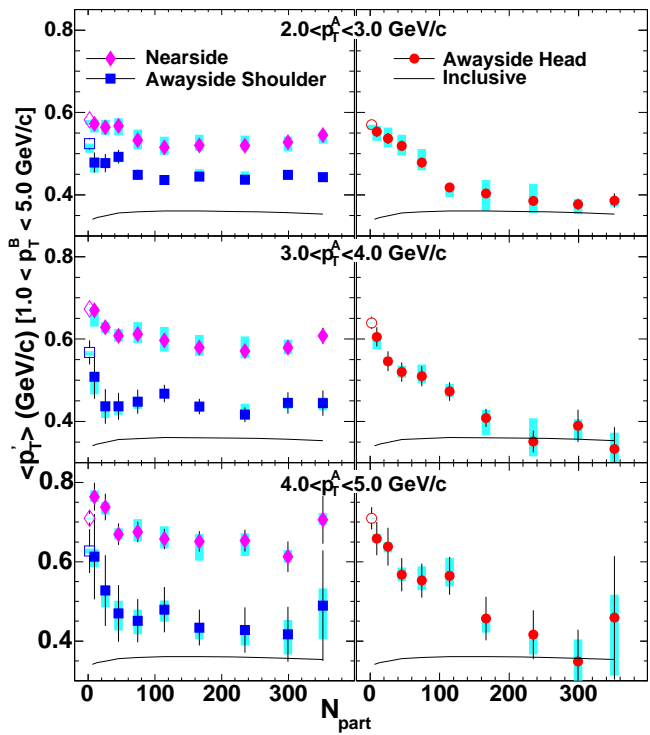}
\caption{\label{fig:slope} Truncated mean $\langle \pt^{\prime}\rangle$ 
in $1<\pt^B<5$ GeV/$c$ versus $N_{\rm{part}}$
for the near-side (diamonds), away-side shoulder (circles) and
head (squares) regions for Au+Au (filled) and p+p (open) for three
trigger \pt bins. Solid lines represent measured values for
inclusive charged hadrons~\cite{Adler:2003au}. Error bars
represent the statistical errors. Shaded bars represent the sum of
$N_{\rm{part}}$-correlated elliptic flow and ZYAM error.}
\end{figure}

To further explore the interplay between the HR and the SR, we
focus on the intermediate \pt region, $1<\pt^{B}<5$ GeV/$c$,
where the medium-induced component dominates the away-side yield.
We characterize the inverse local slope of the partner yield in
this \pt range via a truncated mean \pt, $\langle
\pt^{\rm{\prime}}\rangle\equiv\langle \pt^B\rangle|_{1<\pt^B<5
\rm{GeV}/c}$ - 1 GeV/$c$. $\langle \pt^{\rm{\prime}}\rangle$ is
calculated from the jet yields used to make $I_{\rm{AA}}$ in
Fig.~\ref{fig:inte}. Fig.~\ref{fig:slope} shows the
$\langle \pt^{\rm{\prime}}\rangle$ values for the HR, SR and a
near-side region ($|\Delta\phi|<\pi/3$, NR), as a function of the
number of participating nucleons, $N_{\rm{part}}$. The $\langle
\pt^{\rm{\prime}} \rangle$ values for NR have a weak centrality
dependence. Their overall levels for $N_{\rm{part}}>100$ are
$0.533\pm 0.024$, $0.605\pm0.032$ and $0.698\pm0.040$ GeV/$c$ for
the $\pt^A$ ranges 2-3, 3-4 and 4-5 GeV/$c$,
respectively~\cite{ridge}. This finding is consistent with the
dominance of jet fragmentation on the near-side, i.e. a harder
spectrum for partner hadrons is expected for higher \pt trigger
hadrons.

A very weak centrality dependence is observed for the SR for
$N_{\rm{part}}\gtrsim100$. In this case, the values for $\langle
\pt^{\rm{\prime}} \rangle$ are lower ($\approx 0.45$ GeV/$c$)
and do not depend on $\pt^A$. They are, however, larger than the
values measured for inclusive charged hadrons (0.38 GeV/$c$ shown
by solid lines)~\cite{Adler:2003au}. The relatively sharp increase
in $\langle \pt^{\rm{\prime}} \rangle$ for
$N_{\rm{part}}\lesssim100$ may reflect a significant jet
fragmentation contribution in peripheral collisions. In contrast,
the $\langle \pt^{\rm{\prime}} \rangle$ values for the HR show
a gradual decrease with $N_{\rm{part}}$, starting close to that
for the near-side jet, and approaches the value for the inclusive
spectrum for $N_{\rm{part}}\gtrsim 150$.

The different patterns observed for the yields in the HR and SR
suggest a different origin for these yields. The suppression of
the HR yield and the softening of its spectrum are consistent with
a depletion of yield due to jet quenching. The observed HR yield
could be comprised of contributions from ``punch-through'' jets,
radiated gluons and feed-in from the SR. By contrast, the
enhancement of the SR yield for $\pt^{A,B}<4$ GeV/$c$ suggests a
remnant of the lost energy from quenched jets. However, the very
weak dependence on \pt and centrality (for
$N_{\rm{part}}\gtrsim100$) for its peak location and mean \pt
may reflect an intrinsic property of the response of the medium to
the energetic jets. These observations are inconsistent with
simple deflected jet~\cite{Armesto:2004pt,Chiu:2006pu} and
Cherenkov gluon radiation~\cite{Koch:2005sx} models, since both
the deflection/radiation angle and jet spectra slope would depend
on the $\pt^A$ or $\pt^B$. However, these results are consistent
with expectations for ``Mach Shock'' in a near-ideal
hydrodynamical medium
~\cite{Renk:2005si,Casalderrey-Solana:2004qm}, and thus they can
be used to constrain medium transport properties such as speed of
sound and viscosity to entropy ratio.

In conclusion, we have observed strong medium modification of
away-side shapes and yields for jet-induced pairs in Au+Au
collisions at $\sqrt{s_{\rm{NN}}}$=200 GeV. The detailed
dependence of these results on \pt and centrality gives strong
evidence for two distinct contributions from the regions of
$\Delta\phi\sim\pi$ and $\Delta\phi\sim\pi\pm1.1$. The former is
consistent with jet quenching. The latter exhibits \pt and
centrality independent shape and mean \pt, possibly reflecting
an intrinsic property of the medium response to energetic jets.
These results provide strong constraints on competing mechanisms
for the energy transport.

%%%%%%%%%%%%%%%%%%%%%%%%%  Acknowledgements

We thank the staff of the Collider-Accelerator and Physics
Departments at BNL for their vital contributions. We acknowledge
support from the Department of Energy and NSF (U.S.A.), MEXT and
JSPS (Japan), CNPq and FAPESP (Brazil), NSFC (China), MSMT (Czech
Republic), IN2P3/CNRS and CEA (France), BMBF, DAAD, and AvH
(Germany), OTKA (Hungary), DAE (India), ISF (Israel), KRF and
KOSEF (Korea), MES, RAS, and FAAE (Russia), VR and KAW (Sweden),
U.S. CRDF for the FSU, US-Hungarian NSFOTKA- MTA, and US-Israel
BSF.

%%%%%%%%%%%%%%%%%%%%%%%%% References

\end{document}